\documentclass[aps,prl,twocolumn,superscriptaddress,showpacs,floatfix,longbibliography]{revtex4-2}
\usepackage{amsmath,amssymb,amsfonts,float,graphics,epsfig,epstopdf,color,verbatim,tabularx,bm,multirow,appendix,wasysym}
\usepackage{amsthm}
\usepackage[utf8]{inputenc}
\usepackage[T1]{fontenc}
\usepackage{xcolor}
\usepackage{dsfont}
\usepackage{textcomp}
\usepackage{yfonts}
\usepackage{bm}
\usepackage{subfigure}
\usepackage{mathrsfs}
\usepackage{graphicx}
\usepackage{verbatim}
\usepackage{hyperref}
\usepackage{multirow}
\usepackage{braket}
\usepackage[normalem]{ulem}
\usepackage{tikz}
	\usetikzlibrary{calc}
	\usetikzlibrary{shapes.multipart}

\newcommand{\comments}[1]{}

\newcommand{\stkout}[1]
{\ifmmode\text{\sout{\ensuremath{#1}}}\else\sout{#1}\fi}

\usepackage{natbib} 
\begin{document}

\title{Bipartite reweight-annealing algorithm of quantum Monte Carlo to extract large-scale data of entanglement entropy and its derivative}

\author{Zhe Wang}
\affiliation{Department of Physics, School of Science and Research Center for Industries of the Future, Westlake University, Hangzhou 310030,  China}
\affiliation{Institute of Natural Sciences, Westlake Institute for Advanced Study, Hangzhou 310024, China}

\author{Zhiyan Wang}
\affiliation{State Key Laboratory of Surface Physics and Department of Physics, Fudan University, Shanghai 200438, China}
\affiliation{Department of Physics, School of Science and Research Center for Industries of the Future, Westlake University, Hangzhou 310030,  China}

\author{Yi-Ming Ding}
\affiliation{State Key Laboratory of Surface Physics and Department of Physics, Fudan University, Shanghai 200438, China}
\affiliation{Department of Physics, School of Science and Research Center for Industries of the Future, Westlake University, Hangzhou 310030,  China}

\author{Bin-Bin Mao}
\affiliation{School of Foundational Education, University of Health and Rehabilitation Sciences, Qingdao 266000, China}

\author{Zheng Yan}
\email{zhengyan@westlake.edu.cn}
\affiliation{Department of Physics, School of Science and Research Center for Industries of the Future, Westlake University, Hangzhou 310030,  China}
\affiliation{Institute of Natural Sciences, Westlake Institute for Advanced Study, Hangzhou 310024, China}

\begin{abstract}
We propose a quantum Monte Carlo scheme capable of extracting large-scale data of R\'enyi entanglement entropy (EE) with high precision and low technical barrier. Instead of directly computing the ratio of two partition functions within different space-time manifolds, we obtain them separately via a reweight-annealing scheme and connect them from the ratio of a reference point. The incremental process can thus be designed along a path of real physical parameters within this framework, and all intermediates are meaningful EEs corresponding to these parameters.  In a single simulation, we can obtain many multiples ($\sim \beta L^d$, d is the space dimension) of EEs, which has been proven to be powerful for determining phase transition points and critical exponents.
Additionally, we introduce a formula to calculate the derivative of EE without resorting to numerical differentiation from dense EE data. This formula only requires computing the difference of energies in different space-time manifolds. The calculation of EE and its derivative becomes much cheaper and simpler in our scheme. %This approach opens a way to numerically detect novel phases and phase transitions by scanning EE over a wide parameter region in large-scale systems. 
We then demonstrate the feasibility of using EE and its derivative to find phase transition points, critical exponents, and various phases.
\end{abstract}

\date{\today}
\maketitle

\noindent\textbf{Introduction}\\
With the rapid development of quantum information, its intersection with condensed matter physics has been attracting increasing attention in recent decades~\cite{Amico2008entanglement,Laflorencie2016}. 
One important topic is to probe the intrinsic physics of many-body systems using the entanglement entropy (EE)~\cite{vidal2003entanglement,Korepin2004universality,Kitaev2006,Levin2006,d2020entanglement}. For example, among its many intriguing features, it offers a direct connection to the conformal field theory (CFT) and provides a categorical description of the problem under consideration~\cite{Calabrese2008entangle,Fradkin2006entangle,Nussinov2006,Nussinov2009,CASINI2007,JiPRR2019,ji2019categorical,kong2020algebraic,XCWu2020,ding2008block,Tang2020critical,JRZhao2020,XCWu2021,JRZhao2021,BBChen2022,JRZhao2022,YCWang2021DQCdisorder,YCWang2021U1,jiangFermion2022,zyan2021entanglement}. 
Using EE to identify novel phases and critical phenomena represents a cutting-edge area in the field of quantum many-body numerics. A particularly recent issue is the dispute at the deconfined quantum critical point (DQCP)~\cite{senthil2004deconfined,senthil2004quantum,shao2016quantum}. The EE at the DQCP, e.g., in the $J$-$Q$ model~\cite{sandvik2007evidence,lou2009antiferromagnetic}, exhibits markedly different behaviors compared with those in normal criticality within the Landau-Ginzburg-Wilson paradigm~\cite{JRZhao2021,deng2024diagnosing,d2024entanglement,song2023deconfined,song2023extracting,torlai2024corner}. 
According to the prediction from the unitary CFT~\cite{casini2007universal,casini2012positivity}, the EE with a cornered cutting at the criticality should follow the behaviours $s=al-b\mathrm{ln}l+c$, where $s$ is the EE and $l$ is the length of the entangled boundary, in which the coefficient $b$ cannot be negative. 
However, some recent quantum Monte Carlo (QMC) studies show that $b$ is negative, which seemingly suggests that the DQCP in the $J$-$Q$ model is not a unitary CFT, possibly indicating a weakly first-order phase transition~\cite{JRZhao2022,song2023extracting,song2023deconfined}. In contrast, another recent work indicates that the sign of $b$ depends on the cutting form of the entangled region. For a tilted cutting, $b$ is positive and consistent with the emergent SO(5) symmetry at the DQCP~\cite{d2024entanglement}. All in all, the relationship between the EE and condensed matter physics has been growing increasingly closer in recent years.

\begin{figure}[htp]
\centering
\makebox[\textwidth][l]{\includegraphics[width=1.0\columnwidth]{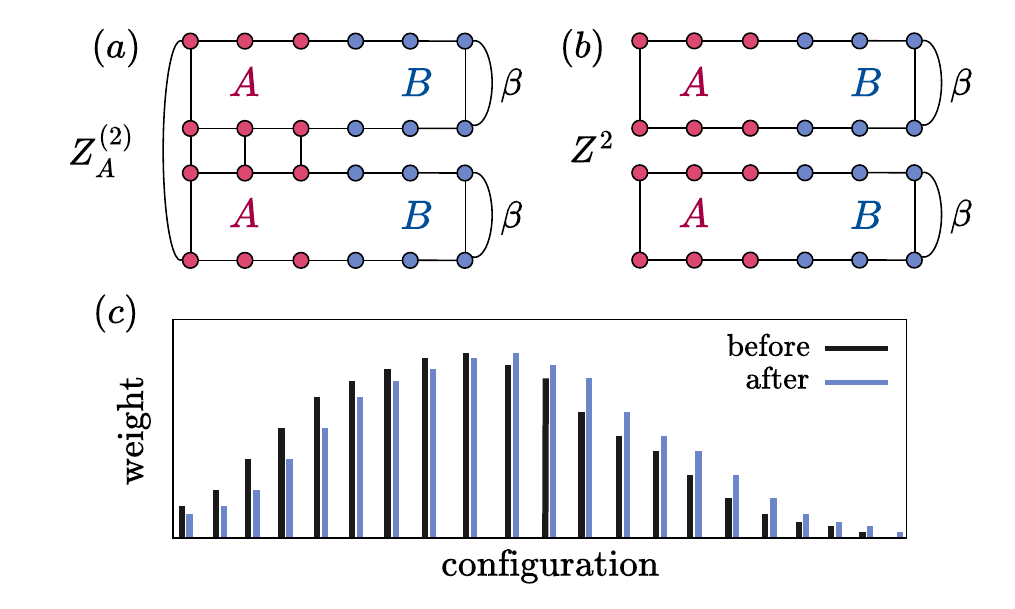}}
\caption{A geometrical presentation of two partition functions within different spacetime manifolds. (a) $Z_{A}^{(2)}=\mathrm{Tr}[\mathrm{Tr}_{B}e^{-\beta H}]^2$ and (b) $Z^2=[\mathrm{Tr}(e^{-\beta H})]^2$, where $H$ is the Hamiltonian of the system. In figure (a), the entangling regions $A$ of two replicas are glued together along the imaginary time direction and the environment regions $B$ of replicas are not connected each other. While the glued region is zero, it becomes back to $Z^2$ as shown in (b). (c) Reweighting a distribution: the sampled distribution (black, before reweighting ) is used to reweight another distribution
(blue, after reweighting), which is reasonable if these two distributions are close to each other as the importance sampling can be approximately kept. }
\label{Fig1}
\end{figure}

However, obtaining high-precision EE via QMC~\cite{Sandvik1999,sandvik2010computational,sandvik2019stochastic,Syljuaasen2002,yan2020improved,suzuki1977monte,PhysRevB.26.5033,suzuki1976relationship,PhysRevE.66.066110,huang2020worm,fan2023clock} with reduced computational cost and a low technical barrier remains a significant challenge in large-scale quantum many-body computations. Although many algorithms have been developed to extract the EE~\cite{hastings2010measuring,humeniuk2012quantum,grover2013entanglement,alba2017out,luitz2014improving,wang2014renyi,herdman2014path,d2020entanglement,song2023resummationbased,zhou2024incremental}, some of which can achieve high precision, the details of these algorithms have become increasingly complex. %Although many algorithms have been developed to extract the EE~\cite{hastings2010measuring,humeniuk2012quantum,grover2013entanglement,alba2017out,luitz2014improving,wang2014renyi,herdman2014path,d2020entanglement,song2023resummationbased,zhou2024incremental}, some of which can also achieve high precision, the details of the algorithm becomes more and more complex. 
Specifically, the $n$th order R\'enyi entropy is defined as $S^{(n)} =\frac{1}{1-n}\ln R_A^{(n)}$.
The key point in extracting the R\'enyi ratio $R_A^{(n)}=Z_A^{(n)}/Z^n$ is to calculate the ratio of two partition functions within different space-time manifolds $Z_A^{(n)}$ and $Z^n$ directly~\cite{hastings2010measuring,luitz2014improving}. In common studies, people usually fix the R\'enyi order $n=2$ as shown in Fig.\ref{Fig1} (a) and (b). 
Due to the area law of EE,  $R_A^{(2)}\propto e^{-al}$ decays to zero rapidly in large systems, where $l$ is the perimeter of the entangled region. Once the ratio $R_A^{(2)}\rightarrow 0$, obtaining high-precision $R_A^{(2)}$ values by QMC based on sampling becomes extremely difficult. 
To overcome this difficulty, the incremental method of the entangled region was introduced~\cite{hastings2010measuring,humeniuk2012quantum}. Its main spirit is one by one adding the lattice sites to increase the entangled region and multiply the ratio of all these intermediate processes to obtain the final ratio. It can be written as: $R_A^{(2)}=Z_A^{(2)}/Z^2=\prod_{i=0}^{N_A-1} Z_{A_{i+1}}^{(2)}/Z_{A_i}^{(2)}$, where the $i$ denotes the number of lattice sites in the entangled region, i.e., $Z_{A_0}^{(2)}=Z^2$ and $Z_{A_{N_A}}^{(2)}=Z_A^{(2)}$. In this way, a super small value has been devided into a product of several larger values. By calculating each intermediate ratio  $Z_{A_{i+1}}^{(2)}/Z_{A_i}^{(2)}$, high-precision $R_A^{(2)}$ can be extracted. The shortcoming of this method is that the number of lattice sites must be an integer, which means the process must be split into a finite number of steps, and some ratios $Z_{A_{i+1}}^{(2)}/Z_{A_i}^{(2)}$ may still be close to zero even after splitting. Moreover, we must note that the replica manifold changes during the calculation due to the intermediate processes in this scheme, which increases the technical barrier of QMC.

To address the finite splitting problem mentioned above, a continuously incremental algorithm of QMC has been developed~\cite{d2020entanglement,alba2017out}. This algorithm involves a virtual process described by a general function  $\tilde{Z}_A^{(2)}(\lambda)$, where $\tilde{Z}_A^{(2)}(\lambda=1)=Z_A^{(2)}$ and $\tilde{Z}_A^{(2)}(\lambda=0)=Z^2$. The problem then becomes calculating the ratio $\tilde{Z}_A^{(2)}(\lambda=1)/\tilde{Z}_A^{(2)}(\lambda=0)$, which can be expressed as $\prod_{\lambda_i} \tilde{Z}_A^{(2)}(\lambda_{i+1})/\tilde{Z}_A^{(2)}(\lambda_i)$. Here  $\lambda$  is a continuous parameter ranging from 0 to 1, thus the interval $[0,1]$ can be divided into any number of segments $\{\lambda_i\}$ according to the computational requirements. This method improves the calculation of EE to unprecedented accuracy and enables the study of systems of unprecedented size. However, the introduction of additional detailed balance (where the entangled region needs being varied during the simulation in this method) imposes specific technical requirements on the code implementation. Moreover, due to the virtually non-physical intermediate processes, the results of these intermediate processes $\tilde{Z}_A^{(2)}(\lambda \neq 1, 0)$ cannot be effectively utilized, leading to waste.

In this paper, we propose a simple method that does not alter the space-time manifold during simulation, and the intermediate process values are physically meaningful and valuable. High-precision EE can now be obtained with lower computational cost and a low technical barrier. Moreover, an efficient scheme for extracting the derivative of EE is proposed for the first time to probe phase transition points.

\vspace{\baselineskip}
\noindent\textbf{Results}\\
\textit{\color{blue}Method.-}
The EE of a subsystem $A$ coupled with an environment $B$ is defined by the reduced density matrix $\rho_A=\mathrm{Tr}_B \rho$, where $\rho=e^{-\beta H}/Z$  and $Z=\mathrm{Tr}e^{-\beta H}$ ($H$ is the Hamiltonian). As mentioned in the introduction, the $n$th order R\'enyi entropy is defined as $S^{(n)}=\frac{1}{1-n}\ln[\mathrm{Tr}(\rho_A^n)]=\frac{1}{1-n}\ln R_A^{(n)}$, where $R_A^{(n)}=Z_A^{(n)}/Z^n$.  The different spacetime manifolds of the two partition functions  $Z_A^{(2)}$ and $Z^2$ (considering $n=2$) are shown in   Fig.\ref{Fig1}.
From the above equations, we know that $Z_A^{(n)}\propto \mathrm{Tr}(\rho_A^n)$ while $Z^n$ is the proportional factor.%\ym{This last sentence may be redundant?}

The normalization factor $Z^n$ is sometimes not important, for example, when we are only concerned with the dynamical information of the entanglement Hamiltonian (e.g., the entanglement spectrum)~\cite{zyan2021entanglement,li2023relevant,wu2023classical,song2023different,liu2023probing,mao2023sampling}. In these cases, only the manifold of $Z_A^{(n)}$ needs to be simulated. However, when we consider the calculation of the EE, the factor becomes non-negligible for obtaining the detailed value. In fact, the hardest difficulty of calculating EE comes from the ratio $R_A^{(n)}=Z_A^{(n)}/Z^n$. This is why the EE algorithms often have to change the manifold between $Z_A^{(n)}$ and $Z^n$.

Unlike the traditional method that directly calculates the ratio $R_A^{(n)}$, we calculate  $Z_A^{(n)}$ and $Z^n$ respectively to avoid the hardness. Let us introduce why we do not need to change the manifold during the simulation. Given a distribution function $Z_A^{(n)}(J)$ (where $Z_A^{(1)}\equiv Z$ without losing generality), and $J$ is a general parameter (e.g., temperature, coupling constants in the Hamiltonian, etc.), the ratio of $Z_A^{(n)}(J')$ and $Z_A^{(n)}(J)$ can be simulated via QMC sampling:
\begin{equation}\label{eq:ratio}
\frac{Z_A^{(n)}(J')}{Z_A^{(n)}(J)} = \bigg\langle \frac{W(J')}{W(J)} \bigg\rangle  _{Z_A^{(n)}(J)}
\end{equation}
where the notation $\langle ...\rangle_{Z_A^{(n)}(J)}$ indicates that the QMC samplings have been performed under the manifold $Z_A^{(n)}$ at parameter $J$. The weights $W(J')$ and $W(J)$ represent the corresponding weights for the same configuration sampled by QMC, but with different parameters $J'$ and $J$ respectively. This means that we simulate the system at parameter $J$ to obtain a set of configurations with weight $W(J)$. Simultaneously, we estimate the corresponding weight $W(J')$ by treating the parameter as $J'$ for the same configuration. The ratio of ${W(J')}/{W(J)}$ can be calculated for each QMC sample to determine the final average, as given in Eq.(\ref{eq:ratio}).

In principle, the ratio ${Z_A^{(n)}(J')}/{Z_A^{(n)}(J)}$ for any $J'$ and $J$ can be solved using the method described above. However, we need to consider how to maintain the importance sampling in our QMC simulation. Clearly, if $J'$ and $J$ are sufficiently close, the weight ratio ${W(J')}/{W(J)}$ is close to 1, making it easier to estimate by sampling. The QMC simulation would be inefficient when the ratio becomes too small or too large. As shown in Fig.\ref{Fig1} (c), if we want to use a known distribution $Z_A^{(n)}(J)=\sum W(J)$ to calculate another distribution $Z_A^{(n)}(J')=\sum W(J')$ by resetting the weight of the samplings, the weights before and after resetting for the same configuration should be close to each other. In this sense, it remains an importance sampling when $J'$ and $J$ are sufficiently close~\cite{troyer2004histogram,ding2024reweightannealing,dai2024residual,neal2001annealed}. Therefore, we introduce the continuously incremental trick to address the issue:
\begin{equation}\label{eq:ratio2}
\frac{Z_A^{(n)}(J')}{Z_A^{(n)}(J)} = \prod_{i=0}^{N-1} \frac{Z_A^{(n)}(J_{i+1})}{Z_A^{(n)}(J_i)}
\end{equation}
where $J_0=J$ and $J_N=J'$, with other $J_i$ values incrementally between the two. Thus, QMC can maintain importance sampling through this reweight-annealing approach~\cite{ding2024reweightannealing,neal2001annealed}.

In this way, we are able to obtain any ratio ${Z_A^{(n)}(J')}/{Z_A^{(n)}(J)}$ in realistic simulations even when the $J'$ and $J$ are far away. However, it still cannot yet give the solution of $Z_A^{(n)}(J')/Z^n(J')$. The antidote comes from some well-known points. Considering that we have calculated the values of ${Z_A^{(n)}(J')}/{Z_A^{(n)}(J)}$ and ${Z(J')}/{Z(J)}$ from the method above, the problem $[Z_A^{(n)}(J')/Z^n(J')=?]$ can be addressed through a known reference point $Z_A^{(n)}(J)/Z^n(J)$. A simple reference point is that $Z_A^{(n)}(J)/Z^n(J)=1$ when the ground state is a product state  $|A\rangle\otimes |B\rangle$. A product state is easy to achieve, for example, by adding an external field in a spin Hamiltonian to polarize all the spins. Of course, other known reference points are also acceptable, such as the state at infinite temperature or a point obtained through other numerical methods. 

One might be concerned about how to deal with a Hamiltonian without a product state in its limit of parameters. An easy approach is to reduce the coupling between $A$ and $B$ to $0$, allowing the ground state to become a product state  $|A\rangle\otimes |B\rangle$ (in this case, the EE reduces to the thermal R\'enyi entropy of isolated $A$, as discussed in Supplementary Note 5 ). In fact, the method of connecting to a reference point is varied.  For example, one can anneal the couplings between separated parts solved by exact diagonalization or hand-weaving from zero to target value, then the problem has also been addressed.
In the Supplementary Note 5, we presented an example where the EE is calculated by annealing the system size starting from the EE of a small system that can be exactly diagonalized.

Now the parameter of the incremental process is continuously tunable, different from the non-equilibrium method~\cite{d2020entanglement,alba2017out}, the incremental path of our method becomes physical and meaningful. It can be set as the real parameter path of a concerned Hamiltonian. In other words, under similar computational cost, the previous method obtains a single point of EE, while ours gains a curve of EEs. A lot of EEs can be obtained in a single simulation, as the number of iterations in the incremental process scales as $\sim \beta L^d$ ($d$ is the space dimension, details are in the Supplementary Note 2 and 3). We will demonstrate that the method is useful for determining the critical points and critical exponents by scanning the EE (see the following section)~\footnote{It is worth noting that if the goal is to capture unknown phase transitions by scanning the EE without requiring the exact value, the number of iterations can be significantly reduced according to your needs}. 

Additionally, we derived a formula to calculate the derivative of the EE (see Eq.(\ref{derivative}) in the following section), which does not require  numerical differentiation from the dense data of EEs and is as simple as computing the fluctuation of energies in different spacetime manifolds. 
The scheme introduced above does not rely on specific detailed QMC methods and particular many-body models. To further understand and test its performance, we will use the spin-1/2 dimerized antiferromagnetic (AFM) Heisenberg model~\cite{Matsumoto2001,ding2018engineering} as an example in the followings. We will use the stochastic series expansion (SSE) QMC method, which we are familiar with, to analyze the model~\cite{Sandvik1999,sandvik2010computational,sandvik2019stochastic,Syljuaasen2002,yan2020improved,yan2019sweeping}.
\begin{figure}[htp]
\centering
\includegraphics[width=\columnwidth]{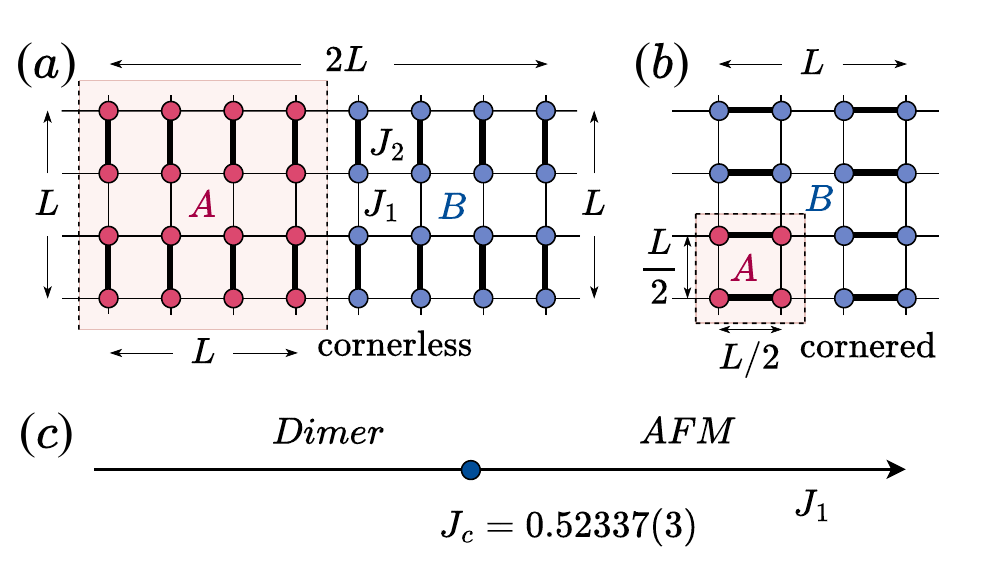}
\caption{Spin-1/2 dimerized AFM Heisenberg model on 2D lattices. The strong bonds $J_2>0$ are indicated by thick lines. The weak bonds $J_1>0$ are indicated by thin lines. (a) The entanglement region $A$ is considered as a $L\times L$ cylinder on  the $2L\times L$ torus with smooth boundaries and with the length  of the entangling region $l=2L$. (b) The entanglement region $A$ is chosen to be a $\frac{L}{2}$ $\times$ $\frac{L}{2}$ square with four corners and boundary
length is $l=2L$. (c) The diagram of the model setting strong bonds $J_2=1$ in which quantum critical point (QCP) is $ J_1=J_c=0.52337(3)$ \cite{Matsumoto2001}. }
\label{Fig2}
\end{figure}

%(a)Square lattice $J_1$-$J_2$ AFM Heisenberg model.The coupling strength of thin bond is $J_1$ while the thick bond represents $J_2$. (b) The phase diagram of the $J_1$-$J_2$ AFM Heisenberg model at $T=0$. (c) The EE converges with increasing the number of intermediate processes under fixed Monte Carlo steps per process.

\textit{\color{blue} Dimerized Heisenberg model.-}
We simulate a spin-1/2 dimerized AFM Heisenberg model on two-dimensional (2D) square lattice as an example to obtain its EE. The Hamiltonian is given by 
\begin{equation}
H=J_1\sum_{\langle ij \rangle}S_iS_j+J_2\sum_{\langle ij \rangle}S_iS_j
\end{equation}
where $\langle ij \rangle$ denotes the nearest-neighbor bonds; $J_1$ and $J_2$ are the coupling strengths of the thin and thick bonds, respectively, as shown in Fig.\ref{Fig2}. Its ground-state phase diagram (Fig.\ref{Fig2} (c)) has been accurately determined by previous QMC studies \cite{Matsumoto2001,ding2018engineering} where the inverse temperature $\beta=2L$ is sufficient to achieve the desired data quality with high efficiency. In the following simulations, we fix $J_2=1$ and tune $J_1$ from $0_+$ to $1$. It is worth noting that the ground state is a dimer product state when $J_1\rightarrow 0$, where the $Z_A^{(n)}/Z^n=1$ if the dimers are not cut by the entangled edge.

In the SSE framework, the Eq.(\ref{eq:ratio}) becomes \cite{d2023leeyang}  
%$\langle(J_1'/J_1)^{n_{J_1}}\rangle_{{Z_A^{(n)}(J_1)}}$, where the $n_{J_1}$ is the number of $J_1$ operators in the SSE sampling, no matter which space-time manifold $Z_A^{(n)}$ is simulated. Thus we have 
\begin{equation}\label{eq:ratio2}
\frac{Z_A^{(n)}(J_1')}{Z_A^{(n)}(J_1)} = \bigg\langle \bigg(\frac{J_1'}{J_1}\bigg)^{n_{J_1}} \bigg\rangle  _{Z_A^{(n)}(J_1)}
\end{equation}
where $n_{J_1}$ is the number of $J_1$ operators in the SSE sampling, regardless of whether the spacetime manifold $Z_A^{(n)}$ or $Z_A^{(1)}\equiv Z$ being simulated \cite{ding2024reweightannealing}. The details of this equation can be found in Supplementary Note 2. 

In the realistic simulation, we need to calculate ${Z_A^{(2)}(J_1')}/{Z_A^{(2)}(J_1=0_+)}$ and ${Z(J_1')}/{Z(J_1=0_+)}$ respectively. We then obtain the final ratio ${[Z_A^{(2)}}(J_1')/Z^2(J_1')]$ based on $[{Z_A^{(2)}}(0_+)/Z^2(0_+)]=1$. 

\begin{figure*}[htp]
\centering
\includegraphics[width=1.0\textwidth]{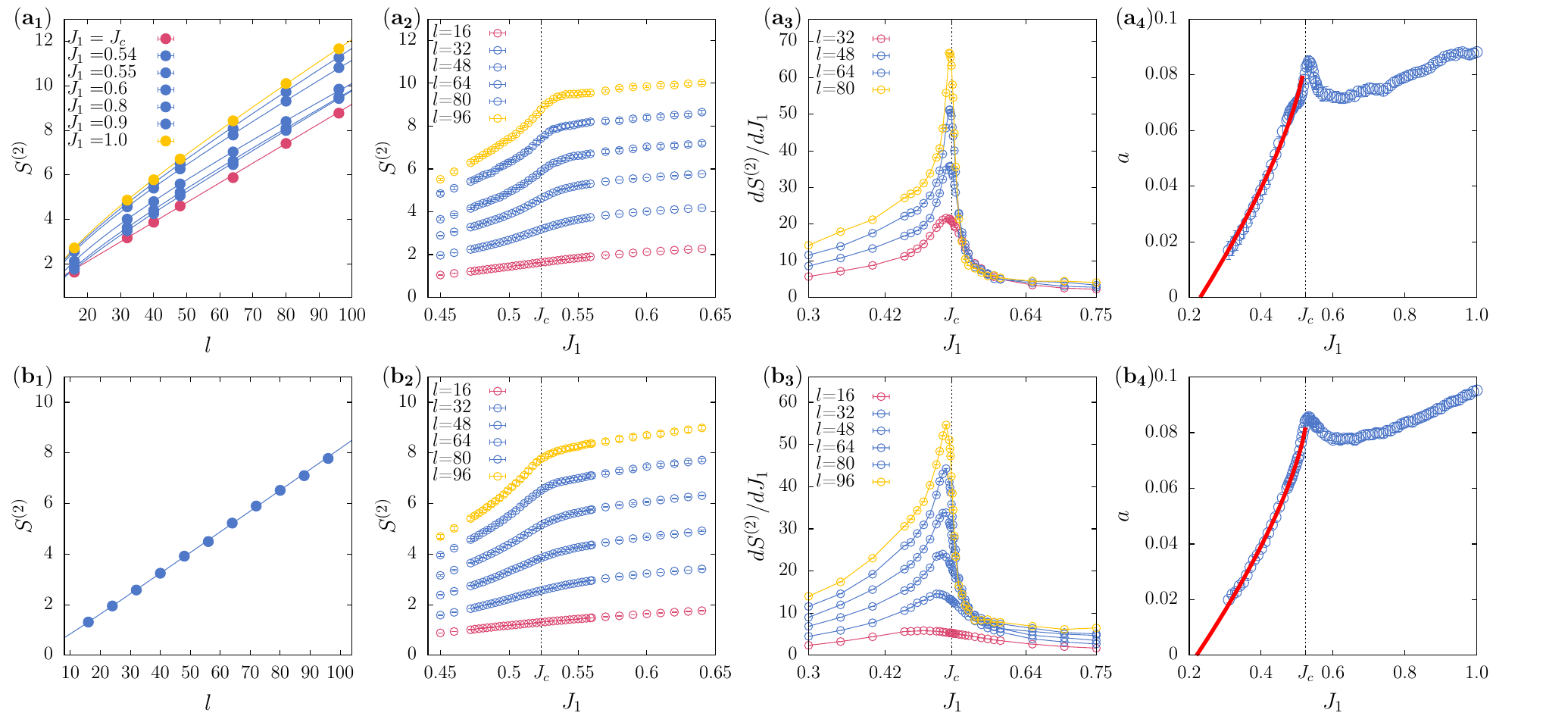}
\caption{2nd Rényi entanglement entropy $S^{(2)}$  of the spin-1/2 dimerized Heisenberg model when the entanglement region $A$ is cornerless [$(a_1)$,$(a_2)$, $(a_3)$ and $(a_4)$ ]  or cornered [$(b_1)$,$(b_2)$, $(b_3)$ and $(b_4)$]. The cornerless/cornered cutting is shown in Fig.\ref{Fig2} (a)/(b). ($a_1$)  The relation between $S^{(2)}$ and entangled perimeter $l$ under different couplings $J_1$. The fitting results are listed in Table \ref{ext1}. ($b_1$) $S^{(2)}$ versus $l$ at the QCP $J_1=J_c=0.52337$. The fitting result is $S^{(2)}(l)=0.083(1)l-0.08(1)\mathrm{ln}l+0.19(2)$ with R/P-$\chi^2$ are 0.85/0.56. [($a_2$) and ($b_2$) ] Scanning $S^{(2)}$ along couplings $J_1$ of different $l$ to identify the critical point. [($a_3$) and ($b_3$) ] The derivative of $S^{(2)}$, $dS^{(2)}/dJ_1$ goes with the coupling $J_1$ in different $l$. The peaks of $dS^{(2)}/dJ_1$ appear at the QCP $J_c$. [($a_4$) and ($b_4$) ] Area law prefactor $a$ versus $J_1$. The red curve is the fitting of $|a(J)-a(J_c)|\sim |J-J_c|^{\nu}$. } 
\label{fig:ee}
\end{figure*}

\begin{ruledtabular}
\begin{table}[!h]
\caption{Fitting results for the data in Fig.\ref{fig:ee} ($\mathrm{a}_1$) with $S^{(2)}(l)=al-b\mathrm{ln}l+c$.	 Reduced and p-value of $\chi^2$ (R/P-$\chi^2$) are also listed. }
\begin{tabular}{l c c c  c }
 	   	   $J_{1}$ 	  & $a$ 	  & $-b$ 	&$-c$  &R/P-$\chi^2$\\
 	   	   	
\hline
	$1.0$			& 0.089(2)   	&1.05(4)     &1.61(9)  &1.00/0.40 \\   
	$0.9$			& 0.085(2)  	&1.02(3)     &1.54(7)  &0.54/0.71  \\
    $0.8$			& 0.079(2)   	&1.06(5)     &1.6(1)   &1.55/0.19 \\
	$0.6$			& 0.072(2)   	&1.06(5)     &2.0(2)   &1.93/0.10 \\
    $0.55$			& 0.078(3)   	&0.8(1)      &1.6(2)   &3.16/0.02 \\
	$0.54$			& 0.08(1)   	&0.6(1)      &1.2(2)   &2.49/0.04 \\				
    $J_c=0.52337$			& 0.08(1)   	    &0.15(17)    &0.1(5)   &2.02/0.1  \\	
\end{tabular}
\label{ext1}
\end{table}
\end{ruledtabular}

\textit{\color{blue} Cornerless cutting.-}
Firstly, we calculate the EE with cornerless cutting as shown in Fig.\ref{Fig2} (a). According to previous works~\cite{JRZhao2021,JRZhao2022,YCWang2021DQCdisorder}, only the entangled edge without cutting dimers (thick bonds) gives correct results consistent with CFT predictions. In Fig.~\ref{fig:ee} ($\mathrm{a}_1$), we present several curves of EE data for different values of $J_1$. The fitting data based on area law are shown in Table.~\ref{ext1}. According to theoretical prediction~\cite{metlitski2011entanglement}, $-b=N_G/2=1$ in the N\'eel phase of the spin-1/2 dimerized Heisenberg model, where $N_G$ means the number of Goldstone modes. Our calculations provide consistent results, as shown in Table.~\ref{ext1}  with $-b\sim 1$ at $J_1=1.0, 0.9, 0.8, 0.6$.
In addition, the theoretical calculation~\cite{metlitski2009entanglement} points out that the $-b=0$ at the Wilson-Fisher O(N) quantum criticality of $d\geq 2$ systems. Our result at $J_c$ in the table also supports this prediction.  We further provide a graph of $-b$ as a function of $J_1$ with some discussions in the Supplementary Note 6. We note that recent works \cite{deng2023improved} have found that the finite size effect in the spin-1/2 AFM Heisenberg model is strong, which notably affects the fitting of the parameter $-b=1$, and a good fitting needs some more corrections considering the finite size effect. However, we find that the simple fitting is not bad in our results. The reason may be that the total system we chose is a rectangle, while the region $A$ is a square, whereas they chose a square total system and a rectangular region $A$. Other QMC works with similar cutting choice as ours also obtains $-b\sim 1$ using  the non-equilibrium algorithm, but in larger sizes~\cite{d2020entanglement,JRZhao2022}. 
Our temperature setting $\beta=2L$ may coincidentally help us approach the correct number of Goldstone modes even in smaller sizes. 

Another advantage of our method is the natural ability to obtain the EE for different parameter values, as shown in Fig.~\ref{fig:ee} ($\mathrm{a}_2$). This allows QMC to probe phase transitions by scanning the EE in 2D and higher-dimensional systems, similar to how the density matrix renormalization group (DMRG) does in 1D~\cite{latorre2004,Legeza2006,Chan2008,Ren2012,Laurell2023}. In Fig.~\ref{fig:ee} ($\mathrm{a}_2$), the convexity of the function changes at the critical point, which is more clearly seen in the derivative of the EE (Fig.~\ref{fig:ee} ($\mathrm{a}_3$)). In the following section, we will introduce a much simpler method to calculate the derivative of the EE without an incremental process and show that the peak of the derivative is located at the QCP. It is worth noting that sometimes the original EE function directly probes the phase transition, while other times the derivative does, which will be carefully discussed in our upcoming work~\cite{wang2024probing}.

\textit{\color{blue} Cornered cutting.-}
For the cornered cutting case, the value $b\sim 0.08$ at the (2+1)D O(3) quantum criticality is also known according to previous theoretical and numerical works~\cite{Inglis2013,KallinJS2014,Helmes2014,JRZhao2021,JRZhao2022}. In Fig.~\ref{fig:ee} ($b_1$), the fitting yields a consistent result of $b=0.08(1)$ at $J_c$. Similar to the cornerless case, the EE for $J_1$ also displays a change in the convexity at the QCP, as shown in Fig.~\ref{fig:ee} ($b_2$). Combined with the data of EE's derivative and the fitting of critical exponent  presented in the next sections, we will find that the shape of the entangled region has little effect on extracting the critical point and critical exponent of the system. 

\textit{\color{blue} EE derivative.-}
It has been proved in the Supplementary Note 1 that the derivative of the $n$th R\'enyi EE can be measured in the form:
\begin{equation}
\frac{dS^{(n)}}{dJ}=\frac{1}{1-n}\bigg[-n\beta\bigg\langle \frac{dH}{dJ} \bigg\rangle_{Z_A^{(n)}}+n\beta\bigg\langle \frac{dH}{dJ} \bigg\rangle_Z \bigg]
\label{derivative}
\end{equation}
where $J$ is a general parameter, $n$ is the R\'enyi index, the first average is simulated on the manifold of $Z_A^{(n)}$ and the second is based on $Z$.
Taking the spin-1/2 dimerized Heisenberg model as an example, with fixed $J_2=1$ and $n=2$, and adjustable parameter $J_1$,  the Eq.(\ref{derivative}) becomes: ${dS^{(2)}}/{dJ_1}=2\beta\langle {H_{J_1}}/{J_1} \rangle_{Z_A^{(2)}}-2\beta \langle {H_{J_1}}/{J_1} \rangle_{Z}$  where  ${H_{J_1}}$ denotes the $J_1$ term of the $H$. Since $H$ is a linear function of $J_1$, this transformation is straightforward. In the SSE framework, this measurement is similar to measuring energy, which is very simple. The details can be found in the Supplementary Note 1.

This conclusion inspires us that we do not need to calculate dense data of EE to obtain the derivative numerically. Instead, simulating the average, $2\beta\langle {H_{J_1}}/{J_1} \rangle_{Z_A^{(2)}}-2\beta \langle {H_{J_1}}/{J_1} \rangle_{Z}$, at the $J_1$ value we concerned is sufficient. We found a similar approach has been used in calculating the derivative of R\'enyi negativity with respect to the inverse temperature~\cite{wu2020entanglement}. Using this method, we have calculated how the derivative of EE goes with $J_1$ in Fig.~\ref{fig:ee} ($\mathrm{a}_3$) and Fig.~\ref{fig:ee} ($b_3$), the peaks of EE derivative locate at the QCP.

\begin{figure}[htp]
\centering
\includegraphics[width=\columnwidth]{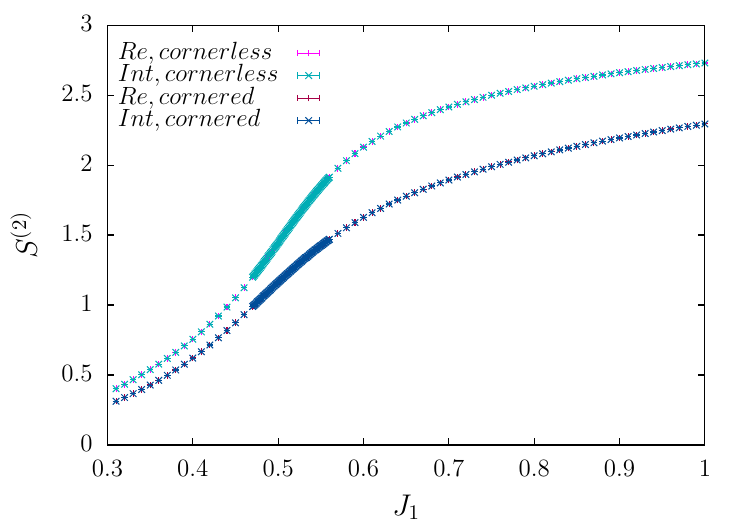}
\caption{2nd Rényi entanglement entropy $S^{(2)}$ of the spin-1/2 dimerized  Heisenberg model as a function of
the coupling $J_1$ are calculated by Reweighting method (Re) and integral method (Int) either in
cornerless or cornered entanglement region $A$ with $l=16$. The results are consistent within errorbar for both methods.}
\label{fig:reandinte}
\end{figure}

In fact, this measurement of EE's derivative also points out another way to calculate the EE through an integral:
\begin{equation}
S^{(n)}(J')=\int_{J_0}^{J'}\frac{dS^{(n)}}{dJ}dJ+S^{(n)}(J_0)
\label{integral}
\end{equation}
where ${dS^{(n)}}/{dJ}$ can be obtained from Eq.~(\ref{derivative}) and the EE $S^{(n)}(J_0)$ at the reference point should be known. We demonstrate the equivalence of the two methods (Eq.~(\ref{eq:ratio2}) and Eq.~\ref{integral}) by taking the spin-1/2 dimerized Heisenberg model as an example, as shown in Fig.~\ref{fig:reandinte}, both in the cornerless and cornered cases. 

We note that Jarzynski’s equality~\cite{PhysRevLett.78.2690} can also be used in our methods, similar to the previous non-equilibrium algorithms~\cite{d2020entanglement,alba2017out}. However, we found that there is almost no acceleration effect for the non-equilibrium version compared with the equilibrium QMC~\footnote{Zhe Wang, Zhiyan Wang, Yi-Ming Ding, Zheng Yan, et al. In preparation}.

\textit{\color{blue} EE and critical behaviors.-} Most previous works have focused on studying the scaling behavior of EE at a known QCP. In this section, we aim to use EE to probe the QCP and extract the critical exponent $\nu$ of a system. We first consider using the parameter position corresponding to the peak of the EE's derivative to determine the QCP of the system. As shown in Fig.~\ref{fig:ee}, it is evident that the peaks of the EE's derivative gradually approach the QCP as the system size increases. We try to obtain the value of the QCP by extrapolating it (see Supplementary Note 7). We find $J_c=0.521(2)$ for  cornerless cutting (dashed line in Fig.~\ref{fig:ee} ($a_3$)) and $J_c=0.521(3)$ for cornered cutting (dashed line in Fig.~\ref{fig:ee} ($b_3$)), which are consistent with the previous result $J_c=0.52337(3)$ within the error bar \cite{Matsumoto2001}. %The details of the extrapolation are provided in the SI.

Performing a fitting of $s=al-b\mathrm{ln}l+c$, we extract in particular the leading area-law coefficient $a$, which is shown in Fig.~\ref{fig:ee} ($a_4$) and ($b_4$) as a function of $J_1$ for both cornered and cornerless cutting. The figures show that $a$ exhibits a non-monotonic behavior as a function of $J_1$ and develops a local maximum at the phase transition point. Similar behavior has been observed in the pioneering work~\cite{Helmes2014}, but in which the normal QMC algorithm costed much more computational resources. The behavior of $a$ in the vicinity of $J_c$ follows an algebraic scaling (considering a $(2+1)$D O(N) QCP): $|a(J)-a(J_c)|\sim |J-J_c|^{\nu}$, where  $\nu$ is the correlation length exponent~\cite{metlitski2009entanglement,Helmes2014}. We are now using this algebraic scaling to extract the critical exponent. 

Let us first consider the case without corners. Setting $J_c$ and $\nu$ as free fitting parameters, we found that $J_c=0.53(1)$ and $\nu=0.88(9)$. The value $0.53(1)$ is consistent with $0.52337(3)$, while $\nu=0.88(9)$ is slightly larger than the $(2+1)$D O(3) universality class $\nu=0.710(2)$ \cite{Matsumoto2001}. 
We then fix $J_c=0.521$ obtained from the EE's derivative above and found $\nu=0.708(31)$, which is consistent with $\nu=0.710(2)$. For the cornered case, we found $J_c=0.526(2)$ and $\nu=0.701(16)$ when setting $J_c$ and $\nu$ as free fitting parameters, which are consistent with $J_c=0.52337(3)$ and $\nu=0.710(2)$ within error bars.  Note that the above fits are all based on the data smaller than the QCP ($J_1<J_c$), because the data larger than the QCP are non-monotonic and difficultly give meaningful results through fitting. Using the known QCP and critical exponent, previous work has already validated $|a(J)-a(J_c)|\sim |J-J_c|^{\nu}$ \cite{Helmes2014}. 
Our method, which naturally generates dense data, can be used to effectively extract the QCP and critical exponent.

%Our results show that the EE is a useful information-theoretic measure of quantum critical phenomena. It does not require a prior knowledge of the quantum phase transitions and should be particularly suitable for detecting exotic QCPs beyond the paradigm of spontaneous symmetry breaking.
%In addition, for the cornered and cornerless cases, we obtained $\gamma=-0.195(6)$ and $\gamma=-0.20(1)$ \zyan{$\gamma$ is c or -c in our scaling equation?} which are consistent within error bars and are in accord with general expectations $\gamma<0$~\cite{Helmes2014,casini2012Renormalization}.

\vspace{\baselineskip}
\noindent\textbf{Discussion}\\
Overall, we develop a practical and unbiased scheme with low technical barrier to extract the high-precision EE and its derivative from the QMC simulations. The spacetime manifold does not need to be changed during the simulation, and the measurement is a simple diagonal observable.  The quantities obtained from intermediate measurements are physical, which makes it possible for QMC to probe novel phases and phase transitions by scanning the EE over large-scale systems in a wide parameter region.

Taking the spin-1/2 dimerized Heisenberg model as an example and scanning along the path from the dimerized phase to the N\'eel order, we found that a peak of EE's derivative instead of EE itself arises at the phase transition point. We have successfully extracted the universal coefficient of the sub-leading term of EE both at O(3) criticality and in the continuous-symmetry-breaking phase. 
Our results demonstrate that EE and its derivative are useful information-theoretic measures of quantum phases and criticalities. 
%Due to EE being a global measure, this method has the potential to detect phase transitions beyond the Ginzburg-Landau symmetry breaking theory. 
In addition, our method is not limited to boson QMC, but can also be applied to other QMC approaches, such as the fermion QMC for highly entangled quantum matters~\cite{jiang2024high}.

%\textit{\color{blue} Acknowledgement}

\vspace{\baselineskip}
\noindent{\bf Methods}

We have developed the bipartite reweight-annealing algorithm of quantum Monte Carlo in this work. Details have been explained in the main text.

\vspace{\baselineskip}
\noindent{\bf Data availability}

The data that support the findings of this study are available at \href{https://github.com/ZheWang-WestLake/Bipartite-reweight-annealing}{https://github.com/ZheWang-WestLake/Bipartite-reweight-annealing}

\vspace{\baselineskip}
\noindent{\bf Code availability}

All numerical codes in this paper are available from the authors.

\vspace{\baselineskip}
\noindent{\bf Acknowledgements}
We thank the helpful discussions with Jiarui Zhao, Dong-Xu Liu, Yin Tang, Zehui Deng, Bin-Bin Chen, Yuan Da Liao, and Wei Zhu. ZY acknowledges the collaborations with Yan-Cheng Wang, Z.Y. Meng and Meng Cheng in other related works. Zhe Wang is supported by the China Postdoctoral Science Foundation under Grants No.2024M752898. BBM acknowledge the Natural Science Foundation of Shandong Province, China (Grant No. ZR2024QA194). The work is supported by the Scientific Research Project (No.WU2024B027) and the Start-up Funding of Westlake University. The authors also acknowledge the HPC centre of Westlake University and Beijng PARATERA Tech Co.,Ltd. for providing HPC resources.start-up funding of Westlake University. 

\vspace{\baselineskip}
\noindent{\bf Author Contributions} 
Z.Y. initiated the work and designed the algorithm. Zhe Wang and Zhiyan Wang performed all the computational simulations. Y.M.D. and B.B.M contributed to the analysis of the results. All authors contributed to the manuscript writing. Z.Y. supervised the project.

%\section{COMPETING INTERESTS}
\vspace{\baselineskip}
\noindent{\bf Competing interests}
The authors declare no competing interests.\\

\clearpage
%\newpage
\appendix
\setcounter{equation}{0}
\setcounter{figure}{0}
\renewcommand{\theequation}{S\arabic{equation}}
\renewcommand{\thefigure}{S\arabic{figure}}
\setcounter{page}{1}
\begin{widetext}
\linespread{1.05}
	
\centerline{\bf\Large Supplementary Information}

\section{Supplementary Note 1: Details about the derivative of entanglement entropy}
The R\'enyi entanglement entropy (EE) is defined as 
\begin{equation}
S^{(n)}=\frac{1}{1-n}\mathrm{ln}\frac{Z_A^{(n)}}{Z^n}
\label{renyiee}
\end{equation}
where 
\begin{equation}
Z=\mathrm{Tr}e^{-\beta H}
\label{z}
\end{equation}
and
\begin{equation}
Z_A^{(n)}=\mathrm{Tr}[(\mathrm{Tr}_Be^{-\beta H})^n] 
\label{zan}
\end{equation}
Thus the EE derivative of $J$ is
\begin{equation}
\frac{dS^{(n)}}{dJ}=\frac{1}{1-n}\bigg[\frac{dZ_A^{(n)}/dJ}{Z_A^{(n)}}-n\frac{dZ/dJ}{Z}\bigg]
\label{deri-EE}
\end{equation}
According to the Supplementary Eq.(\ref{z}), we have
\begin{equation}
\frac{dZ}{dJ}=\mathrm{Tr}[-\beta \frac{dH}{dJ} e^{-\beta H}]
\label{deri-z}
\end{equation}
Thus
\begin{equation}
\frac{dZ/dJ}{Z}=-\beta\bigg\langle \frac{dH}{dJ} \bigg\rangle_Z
\label{dzdjz}
\end{equation}
Similarly, because partial trace is a linear operator which is commutative with the derivative operator, we have
\begin{equation}
\begin{split}
\frac{dZ_A^{(n)}/dJ}{Z_A^{(n)}}&=\frac{\mathrm{Tr}[-n\beta (\mathrm{Tr}_Be^{-\beta H})^{n-1}(\mathrm{Tr}_Be^{-\beta H}\frac{dH}{dJ})]}{\mathrm{Tr}[(\mathrm{Tr}_Be^{-\beta H})^n]}\\
&=-n\beta\bigg\langle \frac{dH}{dJ} \bigg\rangle_{Z_A^{(n)}}
\label{dzdjzan}
\end{split}
\end{equation}
Therefore, the EE derivative can be rewritten as
\begin{equation}
\frac{dS^{(n)}}{dJ}=\frac{1}{1-n}\bigg[-n\beta\bigg\langle \frac{dH}{dJ} \bigg\rangle_{Z_A^{(n)}}+n\beta\bigg\langle \frac{dH}{dJ} \bigg\rangle_Z \bigg]
\label{deri-EE2}
\end{equation}

The equations above is general and doesn't depend on detailed quantum Monte Carlo (QMC) methods. Then let us discuss how to calculate them in stochastic series expansion (SSE) simulation.
For convenience, we fix the R\'enyi index $n=2$ and choose the dimerized Heisenberg model in main text as the example for explaining technical details.
The Hamiltonian is 
\begin{equation}
H=J_1\sum_{\langle ij \rangle}S_iS_j+J_2\sum_{\langle ij \rangle}S_iS_j
\end{equation}
In the following, we fix $J_2=1$ and leave $J_1$ as the tunable parameter. Note the Hamiltonian is a linear function of $J_1$, that means $dH/dJ_1=H_{J_1}/J_1$ in which $H_{J_1}$ is the $J_1$ term in Hamiltonian. Then the EE derivative can be simplified as 
\begin{equation}
\frac{dS^{(2)}}{dJ_1}=\bigg[2\beta\bigg\langle \frac{H_{J_1}}{J_1} \bigg\rangle_{Z_A^{(2)}}-2\beta\bigg\langle \frac{H_{J_1}}{J_1} \bigg\rangle_Z \bigg]
\label{deri-EE3}
\end{equation}
In the SSE frame, it is easy to obtain $\langle H \rangle=\langle -n_{op}/\beta \rangle$~\footnote{How to measure the energy in SSE has been carefully explained in Prof. Sandvik's tutorial \url{http://physics.bu.edu/~sandvik/programs/ssebasic/ssebasic.html}}, where $n_{op}$ is the number of the concerned Hamiltonian operators. Thus the Supplementary Eq.~(\ref{deri-EE3}) can be further simplified to
\begin{equation}
\frac{dS^{(2)}}{dJ_1}=\bigg[-\bigg\langle \frac{n_{J_1}}{J_1} \bigg\rangle_{Z_A^{(2)}}+2\bigg\langle \frac{n_{J_1}}{J_1} \bigg\rangle_Z \bigg]
\label{deri-EE4}
\end{equation}
where $n_{J_1}$ means the number of $J_1$ operators including both diagonal and off-diagonal ones. It's worth noting that there is no ``$2$'' anymore in the $\langle ... \rangle_{Z_A^{(2)}}$ term, because $n_{J_1}$ here contains two replicas' operators which has already been doubled actually.

So far, we have explained how to define the EE derivative in general QMC methods and measure it in detailed SSE algorithm. %As shown in the main text, the integral of the EE derivative is highly consistent with the EE original function, which provides another way to obtain EE. The EE derivative also successfully help us to probe quantum phase transition through either conered or cornerless cutting.

\section{Supplementary Note 2: The weight ratio in SSE}
In the SSE, the partition function can be expanded as~\cite{Sandvik1991,Sandvik1999},
\begin{equation}
\begin{split}
Z =& \sum\limits_{\{\alpha_i\}}  {\beta^n(M-n)! \over M!}
 \langle \alpha_1 |H_{12}|\alpha_2\rangle \times \\
&\langle \alpha_2| H_{23}|\alpha_3\rangle ...\langle\alpha_M |H_{M1}| \alpha_1 \rangle\\
=&\sum\limits_{\{\alpha_i\}} W(\{\alpha_i\})
 \end{split}
\label{zm}
\end{equation}
where $n$ is the number of non-identity operators and $M$ is the cut-off number of the expansion. $H_{ij}$ means the operator connects two closest states $\alpha_i$ and $\alpha_j$.

In the reweighting process, for example, if we only tune the parameter $J_1$, the weight ratio under a fixed $\{\alpha_i\}$ will becomes
\begin{equation}
%\begin{split}
\frac{W(J_1')}{W(J_1)}=\bigg(\frac{J_1'}{J_1}\bigg)^{n_{J_1}}
% \end{split}
\label{wratio}
\end{equation}
where the $n_{J_1}$ is the number of $J_1$ operators. The Supplementary Eq. (\ref{wratio}) comes from the Supplementary Eq. (\ref{zm}), because only the elements $\langle \alpha_i| H_{ij}|\alpha_j\rangle$ in which the $H_{ij}$ is $J_1$ term will affect the ratio.

Similarly, we can get the weight ratio in a general partition function $Z_A^{(n)}$. In fact, its result is also the Supplementary Eq. (\ref{wratio}).

\begin{figure}[htp]
\centering
\includegraphics[width=0.5\textwidth]{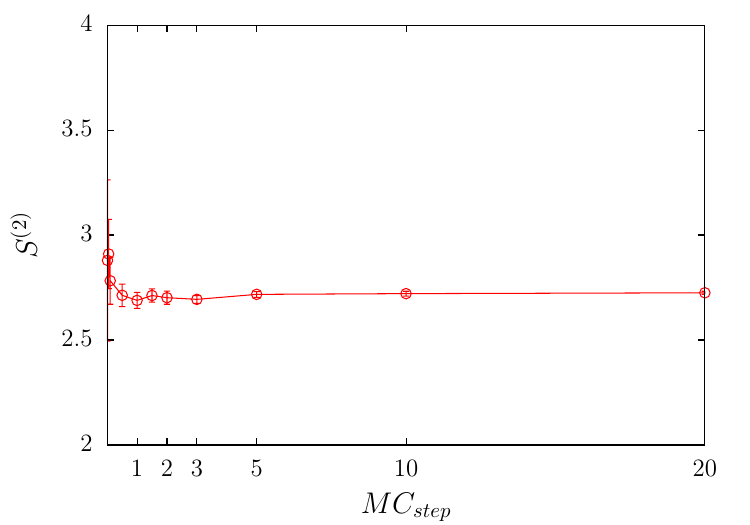}
\caption{Convergence of EE for Monte Carlo step ($MC_{step}$). 2nd Rényi entanglement entropy $S^{(2)}$ of the spin-1/2 dimerized Heisenberg model at $J_1/J_2=1.0$ for $l=16$. The unit of the horizontal axis is $10^3$.}
\label{fig:converge}
\end{figure}

\section{Supplementary Note 3: Details about calculating $R_A^{(n)}$}
As mentioned in the main text, in principle, if $Z_A^{(n)}(J_{1(i+1)})/Z_A^{(n)}(J_{1(i)})$ tends to 1, the calculation results will be more accurate according to the importance sampling, but it requires more segmentation. %Considering that   numerical studies generally require certain data accuracy, we can save CPU time by setting reasonable $Z_A^{(n)}(J_{1(i+1)})/Z_A^{(n)}(J_{1(i)})=0.8$ or $0.5,0.1...$. 
Moderately, we can choose a value which is not too small or large, near 1.
In this paper, we set the smallest value of $Z_A^{(n)}(J_{1(i+1)})/Z_A^{(n)}(J_{1(i)})$  $= C$ ($C\simeq 0.2$ in this paper). For the model we studied, the $n_{J_1}$ will increase with $J_1$ as $n_{J_1}= A\beta L^d$ ($d=2$ in two dimensional systems and $A$ is a constant) at $J_1=J_2=1$. Based on this, we can estimate the value of $J_{ratio}=J_{1(i+1)}/J_{1(i)}$ as $J_{ratio}\sim e^{\mathrm{ln}[C]/(A\beta L^d)}$. %For convenience, let us absorb A into $\beta$, i.e. $A\beta$ becomes $\beta$, because in principle we have a infinite $\beta\rightarrow \infty$.
The number of segmentation $N$ can be determined according to the $J_{ratio}$.

\section{Supplementary Note 4: Convergence criteria}
As shown in Supplementary Fig.\ref{fig:converge}, we found that the EE converges when the Monte Carlo step exceeds 500 in each segment. In our simulation, we set the Monte Carlo step to 2000 and averaged over 40 bins.  For the random number generators, we used linear congruence random number generators in our work, which are often applied in general example codes~\cite{liu2024Analysis}.

\begin{figure}[htp]
\centering
\makebox[\textwidth][l]{\includegraphics[width=1.0\columnwidth]{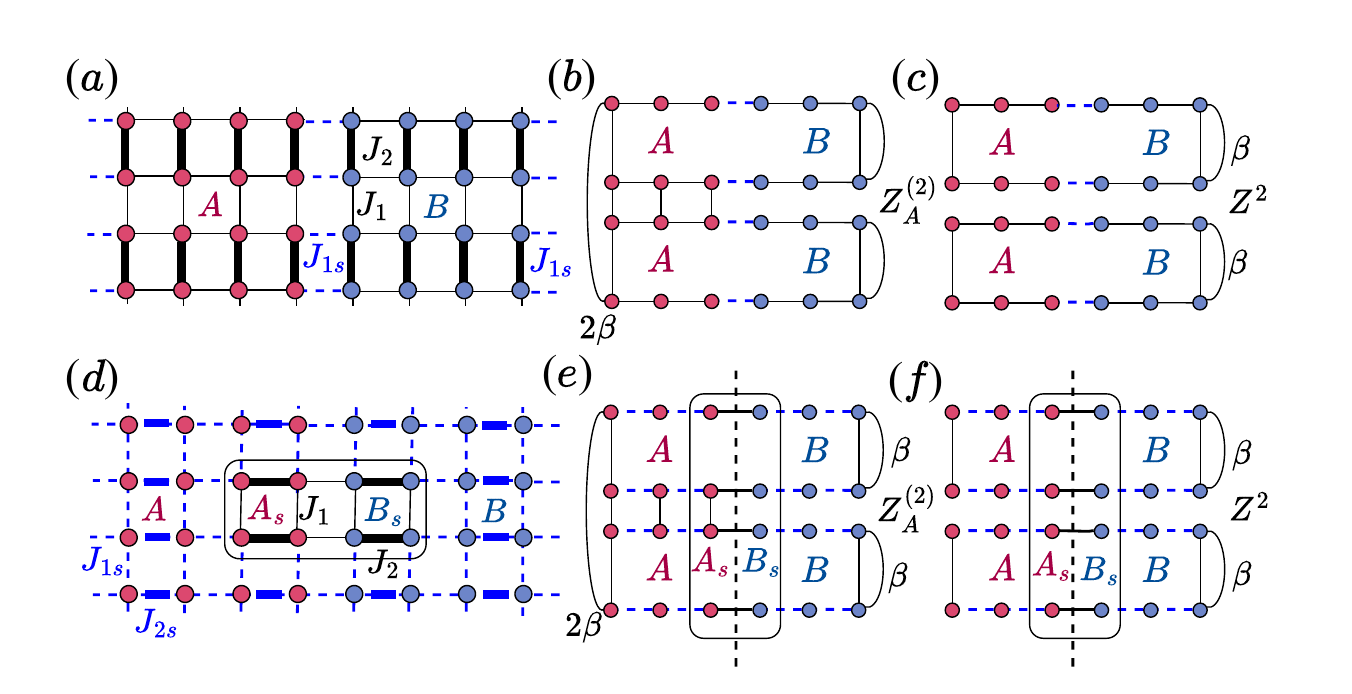}}
\caption{We consider $2$nd order R\'enyi entropy here. (a-c) Annealing edge interaction. (a) The blue dashed line represents the interaction $J_{1s}$ between the entangled region A and the environment region B. During the annealing process, $J_{1s}$ decreases from  $J_{1s}=J_1$ to zero. Geometrical representations of the partition function (b) $Z_{A}^{(2)}$ and (c) $Z^2$. The entangling region $A$ between replicas is glued together in the replica imaginary time direction, while the environment region $B$ for each replica remains independent in the imaginary time direction. %When the glued region is zero, the system returns to $Z^2$. $Z_{A}^{(2)}(J_{1s}=0)=Z_{A}(2\beta, J_1,J_2) \otimes [Z_{B}(\beta, J_1,J_2)]^2$ and $Z^2(J_{1s}=0)=[Z_{A}(\beta,J_1,J_2)]^2\otimes[Z_{B}(\beta,J_1,J_2)]^2$ when $J_{1s}=0$. 
}
\label{sl}
\end{figure}
\section{Supplementary Note 5: Annealing edge interaction or system size}
Here, we consider two other iterative processes to calculate EE. One is to anneal edge interactions $J_{1s}$ (see dashed bonds in Supplementary Fig.\ref{sl} (a-c)) which means decreasing the coupling between the entangled region A and the environment region B from a desired value (here we consider $J_{1s}=J_1$ ) to 0. As mentioned in the main text, if the Hamiltonian has no product states in its limit of parameter, we can anneal the coupling between A and B to obtain a reference point.
The details are as follows: 

\begin{figure}[htp]
\centering
\includegraphics[width=0.5\textwidth]{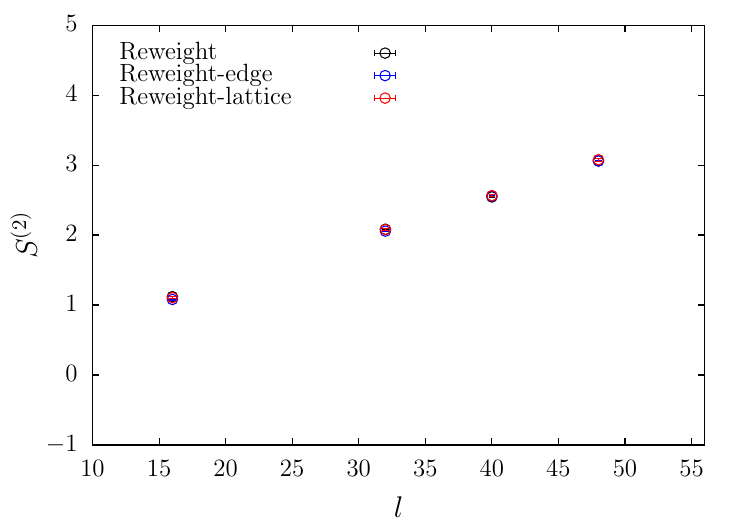}
\caption{Second Rényi entanglement entropy $S^{(2)}$ of the spin-1/2 dimerized Heisenberg model at $J_1/J_2=0.46$ for different $l$ is calculated using three iterative processes.  These processes are: "Reweight" (as described in the main text), "Reweight-edge" (annealing edge interactions), and "Reweight-lattice" (annealing system size). }
\label{fig:threemethod}
\end{figure}

The $n$th order R\'enyi entropy (for convenience, we consider n=2 here) is defined as $S^{(2)}=-\ln[\mathrm{Tr}(\rho_A^2)]
=-\ln (Z_A^{(2)}/Z^2)$. Annealing the edge interactions to $0$, as shown Supplementary Fig.\ref{sl} (a-c), the EE becomes $S^{(2)}=\mathrm{Tr}(e^{-2\beta H_A})[\mathrm{Tr}(e^{-\beta H_B})]^2/\{[\mathrm{Tr}(e^{-\beta H_A})]^2[\mathrm{Tr}(e^{-\beta H_B})]^2\}=\mathrm{Tr}(e^{-2\beta H_A})/[\mathrm{Tr}(e^{-\beta H_A})]^2$ when $J_{1s}=0$, where $H_A$ and $H_B$ are the Hamiltonian of the A and B. Thus the EE becomes the thermal R\'enyi entropy of part A in this case, which represents the ground state degeneracy of part A at zero temperature.

We choose ($J_1=0.46$ and $J_2=1.0$ in dimer phase) as an example to calculate EE for different $l$ by decoupling A and B. The numerical results are shown in Supplementary Fig.\ref{fig:threemethod}, which are consistent with our previous results in the main text.

The other one is to anneal the system size (see Supplementary Fig.\ref{sl} (d-f)) which means that if we obtain the EE for a small system, we can use it as a reference to obtain the EE for a larger system. In Supplementary Fig.~\ref{sl} (d-f)), we provide a schematic diagram to show how to obtain the EE of $8\times 4$ lattice from the EE of $4\times 2$ lattice. The details are as follows (we fix the ratio $J_{1s}/J_{2s}$ during the annealing):

Similar as above, when $J_{1s},~ J_{2s}\rightarrow 0$, the EE becomes 
$Z_A^{(2)}/Z^{2}=(Z_{A_s\cup B_s}^{(2)}/Z_{A_s\cup B_s}^{2})(\prod_{i\in A-A_s} Z_{i}^{(2)}/Z_{i}^{2})$, where $A-A_s$ denotes the sites in the decoupled part $A-A_s$ and $Z_{i}^{(2)}/Z_{i}^{2}=2$ in spin-1/2 systems because the degeneracy is 2 for a free spin. $S_{A_s}^{(2)}$ of a $4\times 2$ lattice can be determined exactly by exact diagonalization. We choose ($J_1=0.46$ and $J_2=1.0$ ) as an example to calculate EE for different $l$. The numerical results are shown in Supplementary Fig.~\ref{fig:threemethod}, which are consistent with the annealing of edge interactions and our previous results in main text. The same idea can be used to construct the EE of a large size by several small sizes' EEs.

\section{Supplementary Note 6: $-b$ as a function of $J_1$ }

 \begin{figure}[htp]
\centering
\includegraphics[width=0.6\textwidth]{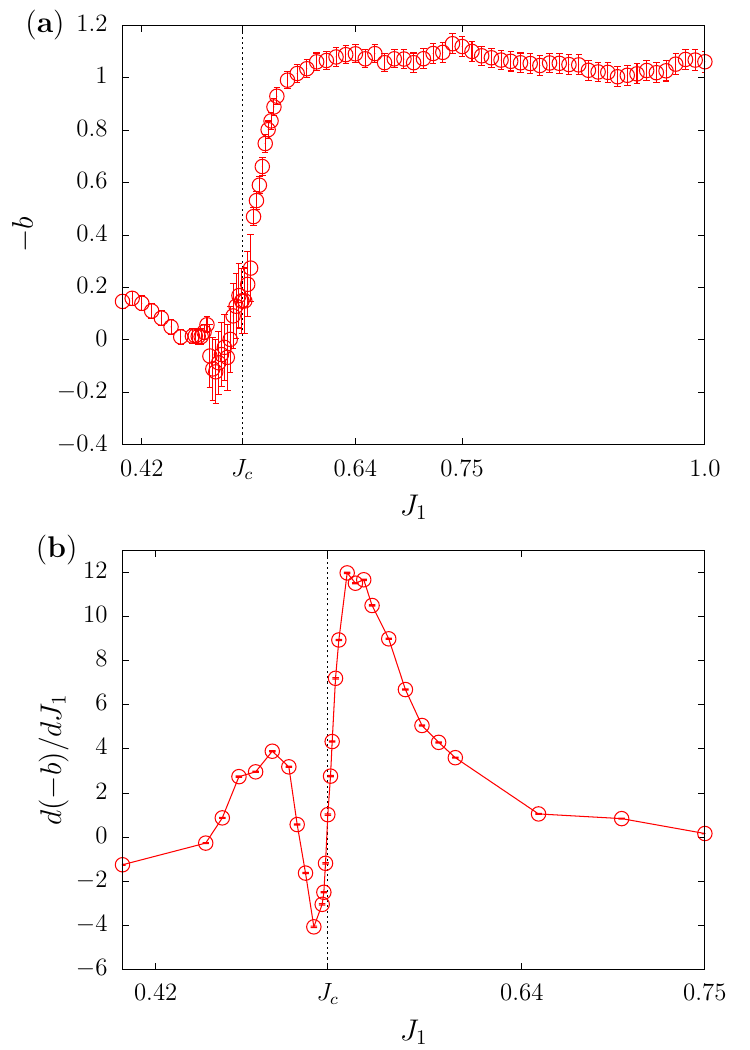}
\caption{(a)$-b$ as a function of $J_1$. (b) $d(-b)/dJ_1$ as a function of $J_1$.}
\label{fig:bln}
\end{figure}

 In Supplementary Fig.~\ref{fig:bln} (a), we plot $-b$ as a function of $J_1$. We find that on the side smaller than and closer to the critical point, $-b$ exhibits a non-monotonic behavior. The $-b$ stays around $1$ in the N\'eel phase and nearly drops to 0 at the phase transition point. 
This behavior is also reflected in its derivative. We also use a same fitting function $al-b\mathrm{ln}l+c$ to fit the derivative of EE which is calculated by Supplementary Eq.(\ref{deri-EE2}). It's worth noting that the calculation of EE's derivative is independent on the simulation of EE, in other words, it is not gained from the numerical derivation of EE data. The EE's derivative in Supplementary Fig.~\ref{fig:bln} (b) is consistent with the curve of EE in (a).

\section{Supplementary Note 7: Extrapolate of peaks  of the EE derivative }
\begin{figure}[htp]
\centering
\includegraphics[width=0.6\textwidth]{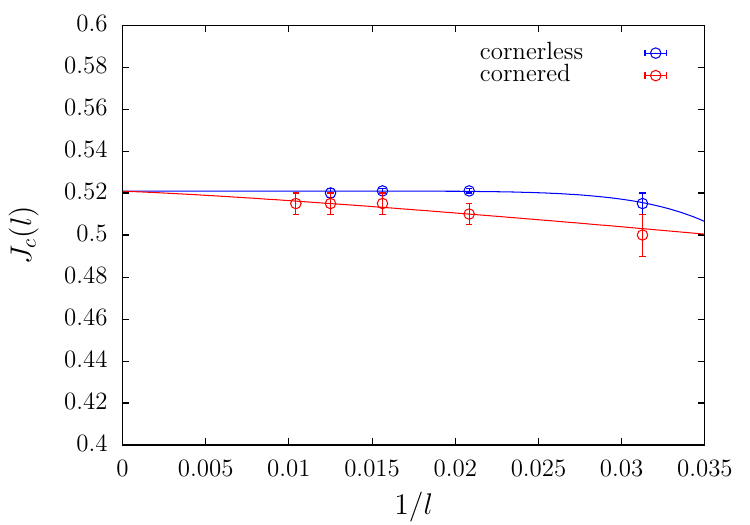}
\caption{Extrapolate of peaks  of the EE derivative for cornerless and cornered cutting. }
\label{fig:deeerxtra}
\end{figure}

As shown in the Fig.3 of main text,  it is evident that the peak of the EE's derivative gradually approaches the QCP of the system as the system size increases. We try to obtain the value of the QCP by extrapolating it. Fitting the data (see Supplementary Fig.~\ref{fig:deeerxtra}) with 
\begin{equation}
    J_c(l)= J_c +b l^{-a},
\end{equation}
we find $J_c=0.521(2)$ for  cornerless cutting  and $J_c=0.521(3)$ for cornered cutting, which are consistent with the previous results $J_c=0.52337(3)$ \cite{Matsumoto2001} within the error bars.

\end{widetext}

\end{document}